\documentclass[preprint,12pt]{emulateapj}

\usepackage{natbib}
\usepackage{verbatim}
\usepackage{graphicx,epstopdf}
\DeclareGraphicsExtensions{.ps}
\usepackage[space]{grffile}
\usepackage{latexsym}
\usepackage{amsfonts,amsmath,amssymb}
\usepackage{url}
\usepackage{longtable}
\usepackage[utf8]{inputenc}
\usepackage{fancyref}
\usepackage{hyperref}
\hypersetup{colorlinks=false,pdfborder={0 0 0},}
\usepackage{textcomp}
\usepackage{multirow,booktabs}
\usepackage{xspace}

\newcommand{\Kepler}{\emph{Kepler}\xspace}
\newcommand{\tess}{\emph{TESS}\xspace}

\usepackage{xcolor}

\definecolor{my_color}{HTML}{3a18b1}

\definecolor{new_color}{HTML}{CF0000}

\shorttitle{Neural network for \tess{}}
\shortauthors{Yu et al.}

\begin{document}

\title{Identifying Exoplanets with Deep Learning III: Automated Triage and Vetting of \tess{} Candidates}

\author{Liang Yu\altaffilmark{1},
Andrew Vanderburg\altaffilmark{2,*},
Chelsea Huang\altaffilmark{1,$\dagger$},
Christopher J.\ Shallue\altaffilmark{3},
Ian J.\ M.\ Crossfield\altaffilmark{1},
B.\ Scott Gaudi\altaffilmark{4},
Tansu Daylan\altaffilmark{1},
Anne Dattilo\altaffilmark{2},
David J.\ Armstrong\altaffilmark{5,6},
George R.\ Ricker\altaffilmark{1}, 
Roland K.\ Vanderspek\altaffilmark{1},
David W.\ Latham\altaffilmark{7}, 
Sara Seager\altaffilmark{1,8,9}, 
Jason Dittmann\altaffilmark{8}, 
John P.\ Doty\altaffilmark{10},
Ana Glidden\altaffilmark{1,8},
Samuel N.\ Quinn\altaffilmark{7}
}
\altaffiltext{1}{Department of Physics, and Kavli Institute for Astrophysics and Space Research, Massachusetts Institute of Technology, Cambridge, MA 02139, USA}
\altaffiltext{2}{Department of Astronomy, The University of Texas at Austin, Austin, TX 78712, USA}
\altaffiltext{3}{Google Brain, 1600 Amphitheatre Parkway, Mountain View, CA 94043, USA}
\altaffiltext{4}{Department of Astronomy, The Ohio State University, Columbus, OH 43210, USA}
\altaffiltext{5}{Centre for Exoplanets and Habitability, University of Warwick, Gibbet Hill Road, Coventry, CV4 7AL, UK}
\altaffiltext{6}{Department of Physics, University of Warwick, Gibbet Hill Road, Coventry, CV4 7AL, UK}
\altaffiltext{7}{Center for Astrophysics \textbar \ Harvard \& Smithsonian, 60 Garden St, Cambridge, MA 02138}
\altaffiltext{8}{Department of Earth and Planetary Sciences, Massachusetts Institute of Technology, Cambridge, MA 02139, USA}
\altaffiltext{9}{Department of Aeronautics and Astronautics, MIT, 77 Massachusetts Avenue, Cambridge, MA 02139, USA}
\altaffiltext{10}{Noqsi Aerospace Ltd, 15 Blanchard Avenue, Billerica, MA 01821, USA}
\altaffiltext{*}{NASA Sagan Fellow}
\altaffiltext{$\dagger$}{Juan Carlos Torres Fellow}

\begin{abstract}
NASA's \emph{Transiting Exoplanet Survey Satellite} (\tess) presents us with an unprecedented volume of space-based photometric observations that must be analyzed in an efficient and unbiased manner. With at least $\sim1,000,000$ new light curves generated every month from full frame images alone, automated planet candidate identification has become an attractive alternative to human vetting. Here we present a deep learning model capable of performing triage and vetting on \tess candidates. Our model is modified from an existing neural network designed to automatically classify \Kepler candidates, and is the first neural network to be trained and tested on real \tess data. In triage mode, our model can distinguish transit-like signals (planet candidates and eclipsing binaries) from stellar variability and instrumental noise with an average precision (the weighted mean of precisions over all classification thresholds) of 97.0\% and an accuracy of 97.4\%. In vetting mode, the model is trained to identify only planet candidates with the help of newly added scientific domain knowledge, and achieves an average precision of 69.3\% and an accuracy of 97.8\%. We apply our model on new data from Sector 6, and present 288 new signals that received the highest scores in triage and vetting and were also identified as planet candidates by human vetters. We also provide a homogeneously classified set of \tess candidates suitable for future training.
\end{abstract}
\bibliographystyle{apj_ads}


\section{Introduction}
The advent of large-scale transit surveys revolutionized our understanding of exoplanets. Both ground-based and space-based telescopes, such as OGLE \citep{udalski}, TrES \citep{alonso}, HATNET/HATS \citep{bakos}, WASP \citep{pollacco}, KELT \citep{siverd2009}, and CoRoT \citep{Auvergne}, have provided us with an unprecedented volume and rate of new discoveries. Perhaps the most notable of all these surveys is NASA's \Kepler space telescope \citep{borucki10, koch10}.
Over the course of its four-year mission, \Kepler observed a total of 200,000 stars, including hosts of more than 2,000 confirmed planets \citep{borucki16}. After the failure of two of its reaction wheels, the repurposed spacecraft \citep[\emph{K2;}][]{howell14} yielded another $\sim 360$ confirmed planets across the ecliptic plane \citep[e.g.][]{crossfield16,mayo18,livingston18a,livingston18b}. \Kepler's successor, the recently launched \emph{Transiting Exoplanet Survey Satellite} \citep[\tess;][]{ricker} will likely more than double the number of known exoplanets \citep{sullivan15, huang18a}. During its two-year mission duration, \tess will observe the sky in $24^{\circ} \times 96^{\circ}$ sectors and downlink data twice during every 27-day sector, eventually covering 20 million stars and 90\% of the sky \citep{sullivan15}. Because \tess observes in the anti-Sun direction \citep{ricker}, \tess targets can be immediately observed from the ground if identified sufficiently rapidly. Prompt follow-up observations are rendered even more crucial by \tess's shorter observing windows, which mean that ephemeris decay (increasing uncertainty in future transit times as we extrapolate our predictions beyond the data used to determine the ephemeris) presents a much bigger problem for \tess than for \Kepler and \emph{K2} (Dragomir et al. in prep.).

Despite the need for rapid and accurate planet candidate identification, space surveys like \tess continue to rely on human vetting.
Typically, teams of experts manually examine possible planet signals and vote on their final dispositions \citep[e.g.][Guerrero et al. in prep]{yu18,crossfield18}.
This process can be quite time-consuming: for a typical \tess sector, it may take a few experienced humans up to a few days to perform triage, i.e. the procedure of rapidly eliminating the obvious false positives, on tens of thousands of candidates. Then, a team of $\sim10$ vetters may spend up to a week classifying the remaining $\sim1,000$ high-quality candidates if we require each one to be viewed by at least three different people. Furthermore, human vetters may not always maintain a consistent set of criteria when judging potential planetary signals. Even an experienced team of vetters may sometimes disagree on the disposition of a TCE, and dispositions given to the same object may vary depending on, for example, the manner of presentation, other TCEs viewed recently, or even the time of day, as we have seen both in \Kepler vetting \citep[e.g.][]{coughlin16} and in our own experience with \tess. 

In response to these shortcomings in human vetting, a number of efforts have emerged to classify light curves automatically and uniformly. Non-machine learning methods make use of classical tree diagrams with criteria designed to mimic the manual process for rejecting false positives \citep{coughlin16,mullally16}. These were completely automated by the end of the \Kepler mission.
Early works on using machine learning to classify \Kepler light curves have explored techniques such as $k$-nearest neighbors \citep{thompson15}, random forests \citep{mccauliff15,mislis16}, and self-organizing maps \citep{armstrong17}. Convolutional neural networks (CNNs), a class of deep neural networks that has proven successful in image recognition and classification, emerged as another possible method.
\citet{zucker18} and \citet{pearson18} investigated the feasibility of using CNNs to detect transiting planets by applying them to simulated data.
The first successful CNN that identified planets in real data from \Kepler was \texttt{AstroNet} \citep{shalluevanderburg}. \citet{ansdell} further improved upon the model by incorporating scientific domain knowledge. Since then, researchers have either modified the original \texttt{AstroNet} model or created their own CNNs to classify candidates from ground-based surveys \citep{schanche19} and \emph{K2} \citep{dattilo}. \citet{osborn19} registered the first attempt to adapt \texttt{AstroNet} for \tess candidates, but the model was trained on simulated data, which are likely to have very different systematics from real \tess data. As a result, the model suffers a deterioration in performance when applied to real \tess data, recovering about 61\% of the previously identified \tess objects of interest.

Here we present the first CNN trained and tested on real \tess data. Our model takes as inputs human-labeled light curves produced by the MIT Quick Look Pipeline (Huang et al. in prep.), and can be trained to perform either triage or vetting on \tess candidates. This paper is organized as follows: In Section~\ref{sec:data}, we describe the creation of the data set used in this study, including how we produced and labeled the light curves; Section~\ref{sec:nn} describes the architecture and training of our neural network for triage and vetting purposes; in Section~\ref{sec:eval}, we evaluate the ability of our neural network to identify planet-like events in the test set; in Section~\ref{sec:new}, we apply our neural network to new data from \tess Sector 6 and present a number of new planet candidates; finally, we discuss some potential improvements to our model in Sector~\ref{sec:future}. All of our code and the list of labeled \tess targets used in this paper are publicly available\footnote{\texttt{AstroNet-Triage:} https://github.com/yuliang419/AstroNet-Triage. \texttt{AstroNet-Vetting}: https://github.com/yuliang419/AstroNet-Vetting. A CSV file containing the list of labeled TCEs used in this study is included in the repositories.}. 

\section{Data set}
\label{sec:data}
Since our goal is to create a neural network capable of performing triage and vetting on \tess light curves, we train and test our models using \tess light curves from Sectors 1-5. Here, we give a brief overview of how these light curves are produced and processed prior to training. We also describe some additional criteria we use to refine this data set. 

\subsection{Identifying Threshold-Crossing Events}
Like \citet{shalluevanderburg}, we work with possible planet signals, which are called ``threshold-crossing events" or TCEs. These are periodic dimming events potentially consistent with signals produced by transiting planets, and are typically identified by an algorithm designed to find such signals. In this study, we adopt the MIT Quick Look Pipeline (QLP; Huang et al. in prep) for light curve production and transit searches. The QLP is partially based on \texttt{fitsh} \citep{pal09}, and is designed to process \tess full-frame images (FFIs) that are obtained with 30-minute time sampling. Immediately upon data downlink, the QLP produces light curves using internal calibrated images from the MIT Payload Operation Center and identifies TCEs. It has already been used to find and alert planet candidates from early \tess sectors \citep[e.g.][]{huang18b, vanderspek18,rodriguez19}. 

\subsubsection{Light Curve Production}\label{sec:lc}

The QLP uses a catalog-based circular aperture photometry method to extract light curves for all stars in the \tess Input Catalog (TIC) with \tess-band magnitudes brighter than 13.5. The apertures are centered based on a predetermined astrometric solution derived on each observed frame using stars with \tess magnitudes between 8-10. The light curves are extracted using five circular apertures. The background is estimated using annuli around the target star on difference images and a photometric reference frame. The photometric reference is computed using the median of 40 frames with minimal scattered light. The difference images are computed using a direct subtraction of the photometric reference frame from the observed frames.     

The light curves produced this way usually contain low-frequency variability from stellar activity or instrumental noise. Following \citet{vj14}, the QLP removes this variability by fitting a B-spline to the light curve and dividing the light curve by the best-fit spline. Outlier points caused by momentum dumps or other instrumental anomalies are masked out prior to detrending. To avoid distorting any transits present, we iteratively fit the spline, remove $3\sigma$ outliers, and refit the spline while interpolating over these outliers \citep[see Fig. 3 in][]{vj14}. We then select an optimal aperture for stars in each magnitude range (13 linear bins between \tess magnitudes of 6-13.5) by determining which aperture size produces the smallest photometric scatter in the magnitude bins.

The light curves are extracted and detrended one \tess orbit at a time, and then stitched together into multi-sector light curves after dividing out the median levels of the detrended light curves.
By Sector 6, stars observed in \tess's continuous viewing zone have light curves with baselines of $\approx$ 166 days, while stars observed in camera 1 (closest to the ecliptic plane) have baselines of only a single $\approx$ 27 day TESS sector. 

\subsubsection{Transit Search}
After producing a detrended light curve for each star using its optimal aperture, the QLP searches the light curves for periodic dipping signals using the Box Least Squares algorithm \citep[BLS;][]{bls}. We perform the search for periods ranging from $0.1$ days, to half the length of the longest baseline expected for the given camera. The number and spacing of frequencies searched by BLS is adapted to the total baseline in the light curves as well, following \citet{vanderburg16}.
We designate any signal with a signal-to-pink-noise ratio \citep[SNR, as defined by][]{vartools} $>9$ and BLS peak significance $>10$ as a TCE. The BLS peak significance is defined as the height of the BLS peak in the spectrum compared to the noise floor of the BLS spectrum.

\subsection{``Ground Truth" Labels}
Unlike the \Kepler DR24 data set used by \citet{shalluevanderburg}, our \tess TCEs do not come with a complete set of human-assigned labels. A small fraction of TCEs underwent group vetting, in which a team of human vetters closely examined the signals using candidate reports created by the QLP and voted on their dispositions, but even this process can yield inconsistent results: a TCE that appears in more than one sector can have different dispositions in different sectors. To ensure homogeneity in the labeling, one of us (LY) visually inspected the light curves of all the TCEs and assigned each to one of four categories: planet candidates (PC), eclipsing binaries (EB), stellar variability (V) and instrumental noise (IS). We used the following set of rules to guide our classification:

\begin{itemize}
    \item Any planet-like signal that does not have a strong secondary eclipse, odd/even transit shape differences, or transit depths that increase with aperture size (indicating that the source of the transit is off-target) is classified as PC.
    \item Some transiting brown dwarfs and M dwarfs have previously been identified as eclipsing binaries in ground-based surveys \citep[e.g.][]{triaud17,collins18} and assigned EB labels in group vetting, but without information beyond the \tess data, even experienced human vetters cannot distinguish these systems from transiting giant planets. We relabel these TCEs as PCs in the data set.
    \item Our data set contains one known planet with visible secondary eclipses, namely the hot Jupiter WASP-18b \citep{hellier09,shporer19}. We assigned this planet to the PC class.
    \item Off-target transit signals whose depths increase with aperture size are always labeled as EBs, regardless of whether the signals could be consistent with planetary transits after correcting for dilution.
    \item Some eclipsing binary systems also exhibit stellar variability. We classify such systems as V if the amplitude of the variability is more than half the eclipse depth, and as EB otherwise.
    \item Any TCEs that are so ambiguous that even human vetters cannot decide whether they are viable planet candidates or false positives are removed from the training set.
    \item PCs and EBs that are significantly distorted by detrending (i.e. if the transits are no longer recognizable as transits, or if their depths change by 50\% or more) are removed from the training set.
    \item We do not make any cuts on transit depth. Deep transit signals that do not show any other signs of being eclipsing binaries are still classified as PCs. The deepest transit in our data set has a depth of 8\%.
    \item Unusual signals that do not fit well into any of the four categories are classified as V. 
\end{itemize}

For the rest of the paper, we assume that these dispositions are the ground truth, even though they may not be perfect. It is likely that a small number of TCEs are misclassified, especially ones that exhibit both stellar variability and eclipses. There may also be a few duplicates in the data set. But since the number of such errors is very small, we expect their impact on our model and performance metrics to be minimal. There are also cases where BLS misidentified the period of a TCE. We corrected as many of these as possible by hand. Occasionally, BLS identifies single-transit events at a fraction of the true period. Our dataset included 20 such singly-transiting EBs and 9 singly-transiting PCs. We do not know the exact periods of these objects, so we use the smallest integer multiple of the BLS period that exceeds the baseline as a guess for the true period. Since the duty cycle of the transit provides information on the density of the host star, which may be useful in distinguishing PCs from EBs (large duty cycles typically indicate that the host star is a giant, and therefore more likely to host EBs), any inaccurate estimates of the period would only be a potential concern in vetting, not in triage. But the number of PCs affected is also small, so again we do not expect them to have a large impact on our model's performance.

After manually assigning labels to all TCEs, we binarize the labels as ``planet-like" and ``non planet-like". When using our neural network to perform triage, both the PC and EB classes are considered to be ``planet-like", so that we retain as many potential planet candidates as possible. When using the network for vetting, we perform a more rigorous selection and only consider PCs as ``planet-like". 

We make use of TCEs from \tess Sectors 1-4, but because the V and IS classes drastically outnumber both PCs and EBs, we supplemented our data set with 296 PCs and EBs from Sector 5. In total, we have 16,516 TCEs for triage, including 493 PCs, 2,155 EBs and 13,868 V and IS combined. If an object is identified as a TCE in multiple sectors, we break up the light curve into individual sectors and count each sector as a separate object. For vetting, another 65 TCEs were discarded due to an insufficient number of points ($<$ 5) to construct secondary eclipse views, resulting in 492 PCs, 2,154 EBs and 13,805 V and IS combined. We randomly shuffle and partition them into three subsets: training (80\%), validation (10\%) and test (10\%). The validation set is used to choose model hyperparameters during training, and the test set to evaluate final model performance.

\subsection{Preparing Input Representations} \label{sec:rep}
Following \citet{shalluevanderburg}, we process each light curve into a standardized input representation before feeding it into the neural network. Since the QLP already removes low-frequency variability from the light curves, we skip the detrending step. The light curve is then phase-folded at the period identified by BLS, such that the transits are lined up and centered. We remove any points corresponding to images with non-zero data quality flags, and any upward outliers that are more than 5 times the median absolute deviation away from the median. 

We then binned the data into two views, similar to those described in \citet{shalluevanderburg}: a ``global view", which shows the light curve over an entire orbital period; and a ``local view", which is a close-up of the transit event, spanning no more than two transit durations on either side of the transit mid-point. \citet{shalluevanderburg} grouped their phase-folded light curves into 2,001 bins for the global view, and 201 bins for the local view. The \Kepler light curves used by \citet{shalluevanderburg} span up to 4 years in duration and contain approximately 70,000 points each. Many \tess light curves, on the other hand, only span about 27 days and have far fewer data points. The resulting phase-folded light curves are therefore much sparser than those from \Kepler. For this reason, we reduced the number of bins in the global and local views to 201 and 61 respectively, and linearly interpolated the data over empty bins. 

In vetting mode, we also prepare a ``secondary eclipse view", which was not present in the original \texttt{AstroNet} model, but was suggested as a possible improvement to the model by \citet{shalluevanderburg}. We first perform a search for the most likely secondary eclipse by masking the transits in the phase-folded light curves and using a BLS-like algorithm to fit a box (whose width is fixed to that of the primary transit) to various positions between orbital phases 0.1 and 0.9 in the masked and folded light curve. The position that yields the highest S/N is assumed to be the midpoint of the most likely secondary eclipse. We then normalize and bin the folded light curve within up to two transit durations on either side of this location into 61 bins, following the exact same procedure we use to produce the local views. 

\begin{figure*}[htb]
    \centering
    \includegraphics[width=\textwidth]{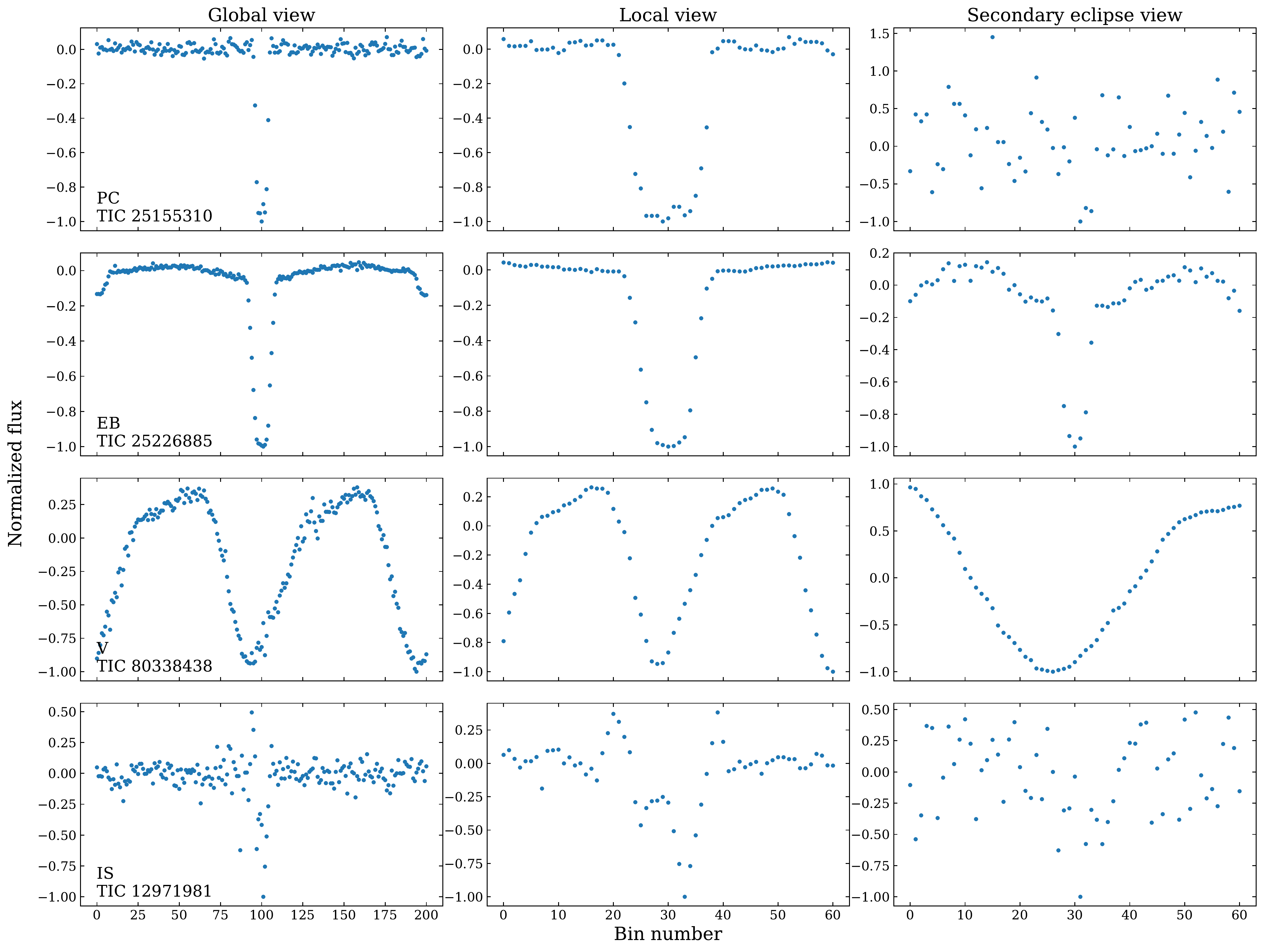}
    \caption{For each TCE, we prepare three phase-folded, depth-normalized representations of the light curve: the ``global view" (left column) is a fixed-length representation of the entire period; the ``local view" (middle column) is a close-up view of the putative transit; the ``secondary eclipse view" (right column), only present in vetting mode, is a close-up view of the most likely secondary eclipse. Each row presents an example from one of the four categories of TCEs: PC (planet candidates), EB (eclipsing binaries), V (stellar variability) and IS (instrumental artifact). }
    \label{fig:LCs}
\end{figure*}

Fig.~\ref{fig:LCs} shows examples of global, local and secondary eclipse views for different classes of signals. 

\section{Neural Network}\label{sec:nn}
\subsection{Architecture}
Our neural network architecture is based on \texttt{AstroNet}, a deep convolutional neural network (CNN) developed by \citet{shalluevanderburg}. CNNs are a class of deep learning model used for inputs with spatial structure (e.g. images or time series).
A CNN contains a hierarchy of ``convolutional layers.'' Each convolutional layer performs a cross-correlation operation by sliding a small filter over the input, summing the result, and adding it to a feature map. Each filter activates in response to a specific feature or pattern in its input. A CNN typically contains many consecutive convolutional layers. In the deeper layers, simpler features learned in previous layers are combined into more complex features. During training, the parameters of the convolutional filters are adjusted to minimize a cost function, a measure of how far the model's predictions are from the true labels in its training set.

\texttt{AstroNet} is implemented in TensorFlow \citep{abadi16}, an open source machine learning framework developed at Google Brain. The global and local view vectors (and secondary eclipse view in vetting mode) are passed through disjoint convolutional columns with max pooling, and then combined in shared fully connected layers ending in a sigmoid activation function. The model outputs a value in $(0,1)$, with values close to 1 indicating high confidence that the input is a transiting planet and values close to 0 indicating high confidence that the input is a false positive. \citet{shalluevanderburg} trained 10 independent copies of the model with different random parameter initializations and averaged the outputs from these 10 copies for all predictions. This technique, known as ``model averaging", improves the robustness of the predictions by averaging over the stochastic differences between the individual models. We refer the interested reader to the \citet{shalluevanderburg} paper for a more detailed description of convolutional neural networks and the associated terminology.

We have made a few key modifications to the original \texttt{AstroNet} architecture, depending on whether the model is used for triage or vetting. Here we describe the two different modes in detail.

\subsubsection{Triage Mode}
The main goal of triage is to eliminate all the obvious non-planetary signals among the TCEs. Most TCEs are caused by instrumental artifacts and stellar variability. The remaining TCEs (usually a mix of planet candidates, eclipsing binaries and blended eclipsing binaries) are then passed on to the vetting stage, where they are examined in more detail. Typically, triage is performed by a human who visually inspects the light curve of each TCE and separates the signals that do not look remotely planet-like at first glance. There are usually a large number of TCEs to be triaged (a few thousand per \tess sector). Our neural network's triage mode, which we dub \texttt{AstroNet-Triage}, is designed to automate the triage process.

\begin{figure}[ht]
    \centering
    \includegraphics[width=0.4\textwidth]{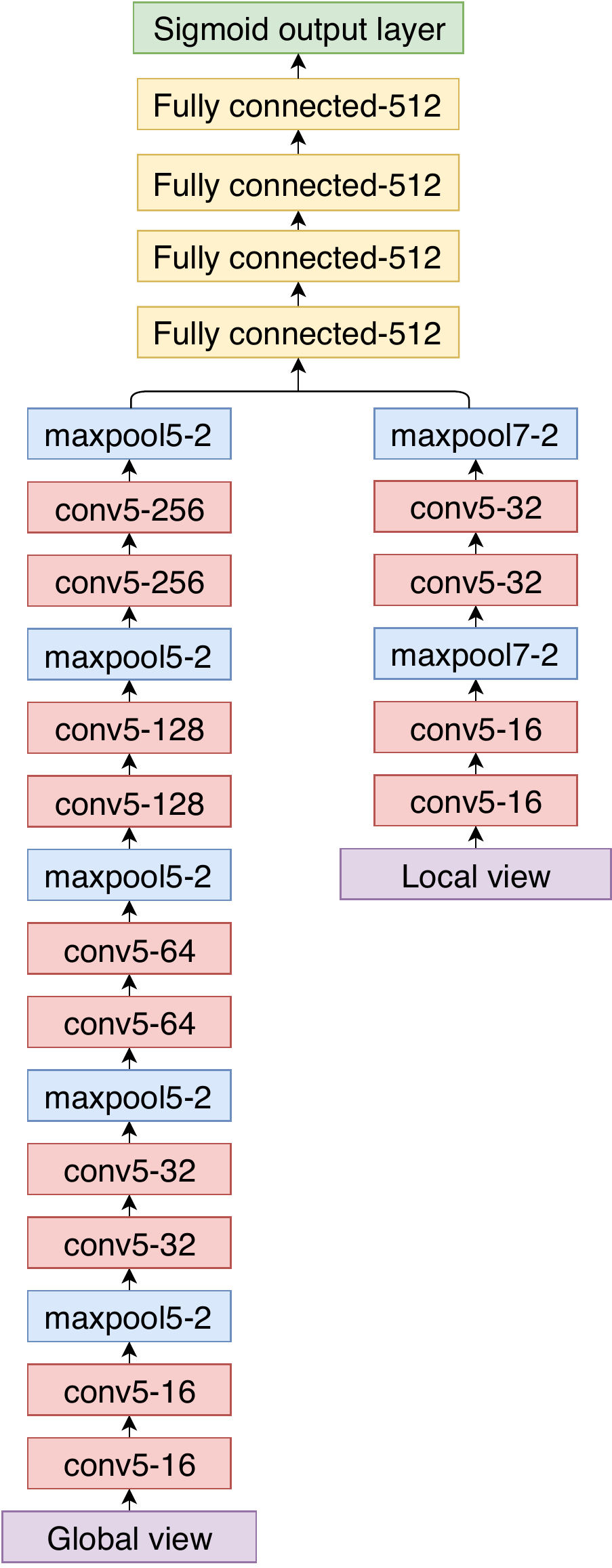}
    \caption{The architecture of \texttt{AstroNet-Triage} \citep[identical to that of the best-performing neural network from][]{shalluevanderburg}. Convolutional layers are denoted conv$<$kernel size$>$-$<$number of feature maps$>$, maxpooling layers are denoted maxpool$<$window length$>$-$<$stride length$>$, and fully connected layers are denoted Fully connected-$<$number of units$>$.}
    \label{fig:triage}
\end{figure}

\texttt{AstroNet-Triage} serves to classify TCEs into ``planet-like" (including PCs and EBs) and everything else. We find that the original \texttt{AstroNet} architecture works well for triage purposes, and that changing the architecture does not yield any significant improvement over the original model, so we make no modifications to the architecture in triage mode. We pass both the global and local views, described in Section~\ref{sec:rep}, through separate convolutional columns before concatenating them in the fully connected layers. We reproduce this architecture in Fig.~\ref{fig:triage}.

\subsubsection{Vetting Mode}
\begin{figure*}[ht]
    \centering
    \includegraphics[width=0.65\textwidth]{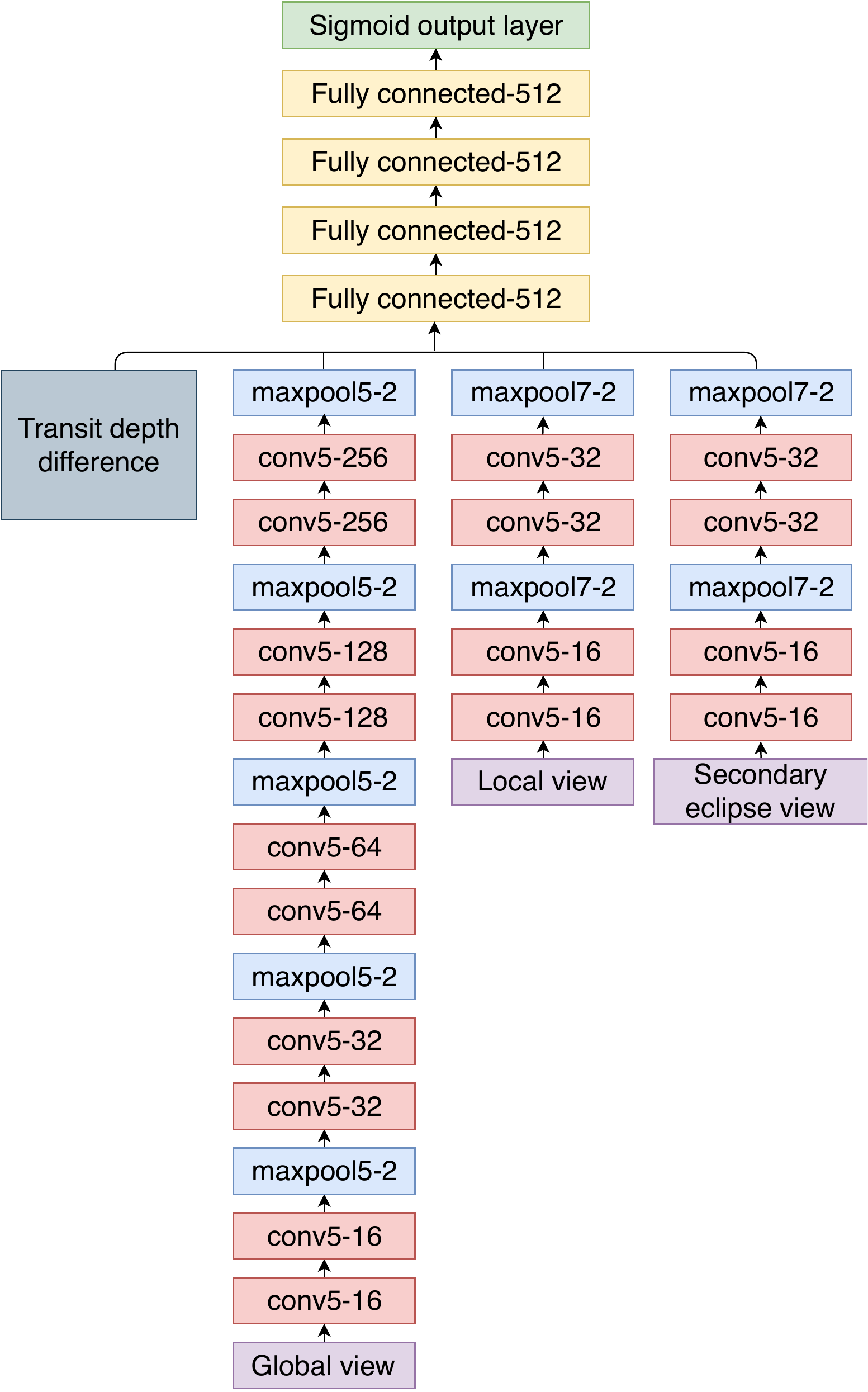}
    \caption{Same as Fig.~\ref{fig:triage}, but for \texttt{AstroNet-Vetting}.}
    \label{fig:vetting}
\end{figure*}

When used for vetting, the model (dubbed \texttt{AstroNet-Vetting}) must also be able to distinguish EBs from PCs. Here we feed the global and local views to the neural network as we do in triage mode, but we also include a close-up of the most likely secondary eclipse (described in Section~\ref{sec:rep}) in a disjoint convolutional column. In addition, we also concatenate a scalar feature to the outputs of the convolutional columns, namely the difference in transit depths measured in two apertures with radii of 2.75 and 3 pixels, divided by the out-of-transit standard deviation measured in the smaller aperture. We chose these two apertures because we find that they are generally large enough to encompass most of the flux from the target star, yet small enough to not include too much flux from background stars. The transit depths are estimated by fitting a box-shaped model to the light curves. 
This ``depth change" feature is normalized by subtracting the mean of the entire training set and dividing by the standard deviation. The motivation behind adding a transit depth difference between different apertures is to help the model identify potential blends. When the source of a transit is off-target, a larger aperture typically produces a deeper transit than a smaller one. Transit depth differences are a simpler alternative to including the entire centroid time series, which \citet{ansdell} and \citet{osborn19} used in their model. Also unlike \citet{ansdell} and \citet{osborn19}, we chose not to incorporate stellar parameters because a substantial fraction of our TCEs simply do not have stellar parameters available. This is because we search all stars in the FFIs, not just those selected for 2-minute-cadence observations. We also experimented with adding the \tess magnitude as a scalar feature, but its effect on the output is negligible.
The architecture of the \texttt{AstroNet-Vetting} model is illustrated in Fig.~\ref{fig:vetting}.

\subsection{Training}
We trained the model for 14,000 steps on the training set in both the triage and vetting modes. We used the Adam optimization algorithm \citep{kingmaba} to minimize the cross-entropy error function over the training set. The number of training steps was chosen to minimize this error function over the validation set. During training, we augmented our training data by applying random horizontal reflections to the light curves with a 50\% probability. This process generates similar but not identical samples with the same labels as the originals, thereby increasing the effective size of our training set and reducing the risk of overfitting. We trained the model with a batch size of 64, a learning rate of $\alpha = 10^{-5}$, and exponential decay rates of $\beta_1 = 0.9$, $\beta_2 = 0.999$ and $\epsilon = 10^{-8}$ \citep[for more details on these parameters, see][]{kingmaba}.

Like \citet{shalluevanderburg}, we also make use of ``model averaging" to improve the robustness of our predictions. We trained 10 independent, randomly initialized copies of the same model and used the average outputs of all copies for all predictions. Each copy may perform better or worse in different regions of parameter space due to its random parameter initialization, but model averaging averages over these differences. It also minimizes the stochastic differences that exist between individual models, thus making different configurations more comparable.

\section{Evaluation of Neural Network Performance}\label{sec:eval}
We assess the performance of our neural network using the test set, the 10\% of TCEs that were randomly left out of the training/validation sets and were thus not used to tune the model or its hyperparameters. Given the highly imbalanced nature of our training set, accuracy - the fraction of TCEs that the model correctly classified - is not a very useful measure of the model's performance, because we can achieve high accuracy simply by classifying everything as negative (not planet-like). The same can be said of the AUC (area under the receiver-operator characteristic curve, equivalent to the probability that a randomly selected positive is assigned a higher prediction than a randomly selected negative). We therefore make use of three additional metrics: precision (reliability), recall (completeness) and average precision. Precision is defined as the fraction of all objects classified as positives that are indeed true positives. Recall is defined as the fraction of all positives in the test set that were correctly classified as positives. There is a trade-off between precision and recall depending on the classification threshold (the score above which we consider an object to be a positive): increasing the threshold typically raises the precision while lowering the recall, and vice versa. Average precision is the weighted mean of precisions achieved at each threshold, with the increase in recall from the previous threshold used as the weight. 

\begin{figure}[htb]
    \centering
    \includegraphics[width=0.5\textwidth]{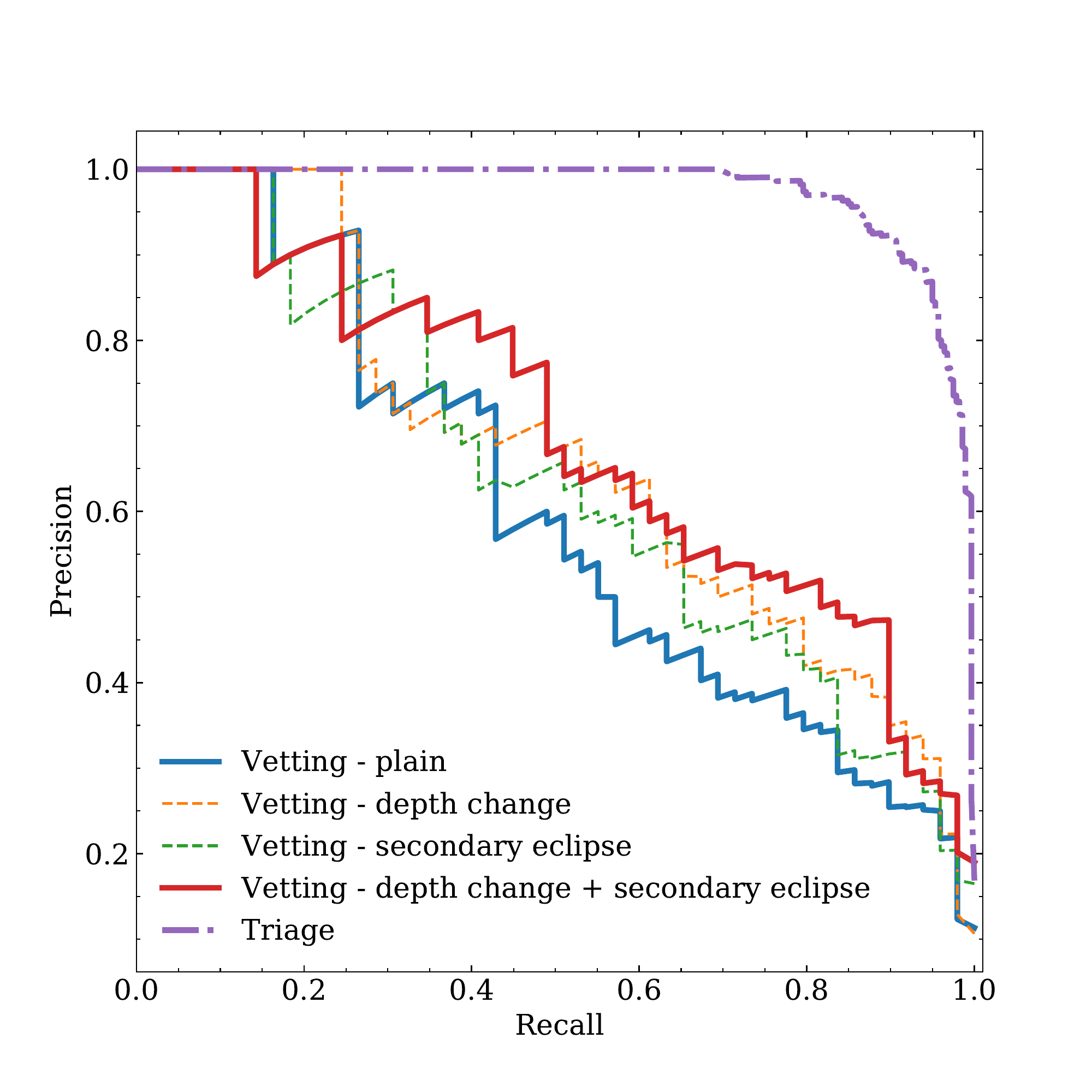}
    \caption{Precision-recall curve of our neural network in both triage and vetting modes. The triage model is trained to distinguish PCs and EBs from obvious false positives, and the vetting model is trained to identify only PCs.
    The line labeled ``vetting - plain" shows the original \texttt{AstroNet} model applied to vetting without the addition of any new features. The two dashed lines show the individual contributions of new features we added: ``vetting - depth change" is the addition of transit depth differences alone, and ``vetting - secondary eclipse" is the addition of secondary eclipse views. ``Vetting - depth change + secondary eclipse" is the final \texttt{AstroNet-Vetting} model that combines both features.}
    \label{fig:pr}
\end{figure}

In Fig.~\ref{fig:pr}, we show the precision-recall (PR) curves for triage and vetting on our test set. Each point on a curve corresponds to the precision and recall values for that model at a different choice of classification threshold. For \texttt{AstroNet-Vetting}, we also plot separate PR curves for the original \texttt{AstroNet} model architecture and models with the two new features added individually to show the impact of each on model performance. Table~\ref{tab:ap} shows the accuracies (calculated for a classification threshold of 0.5), AUC, and average precisions achieved by all of these models on the test set. As mentioned earlier, the models can achieve very high accuracy and AUC in vetting mode and yet still struggle to produce a reliable planet sample.

\begin{table}[htb]
  \caption{Ensembled results achieved on the test set\label{tab:ap}}
  \centering
  \begin{tabular}{lrrr}
\toprule
        Model &  Accuracy    & AUC & Average precision   \\
        \midrule
        Triage & 0.974 & 0.992 & 0.970 \\
        \midrule
Vetting - \texttt{AstroNet} plain & 0.977 & 0.973 & 0.605 \\
Vetting - depth change & 0.978 & 0.980 & 0.669 \\
Vetting - secondary eclipse & 0.976 & 0.978 & 0.642 \\
Vetting - depth change  & 0.978 & 0.984 & 0.693 \\
+ secondary eclipse \\
\bottomrule
  \end{tabular}
\end{table}

\begin{figure*}[hbt]
    \centering
    \includegraphics[width=\textwidth]{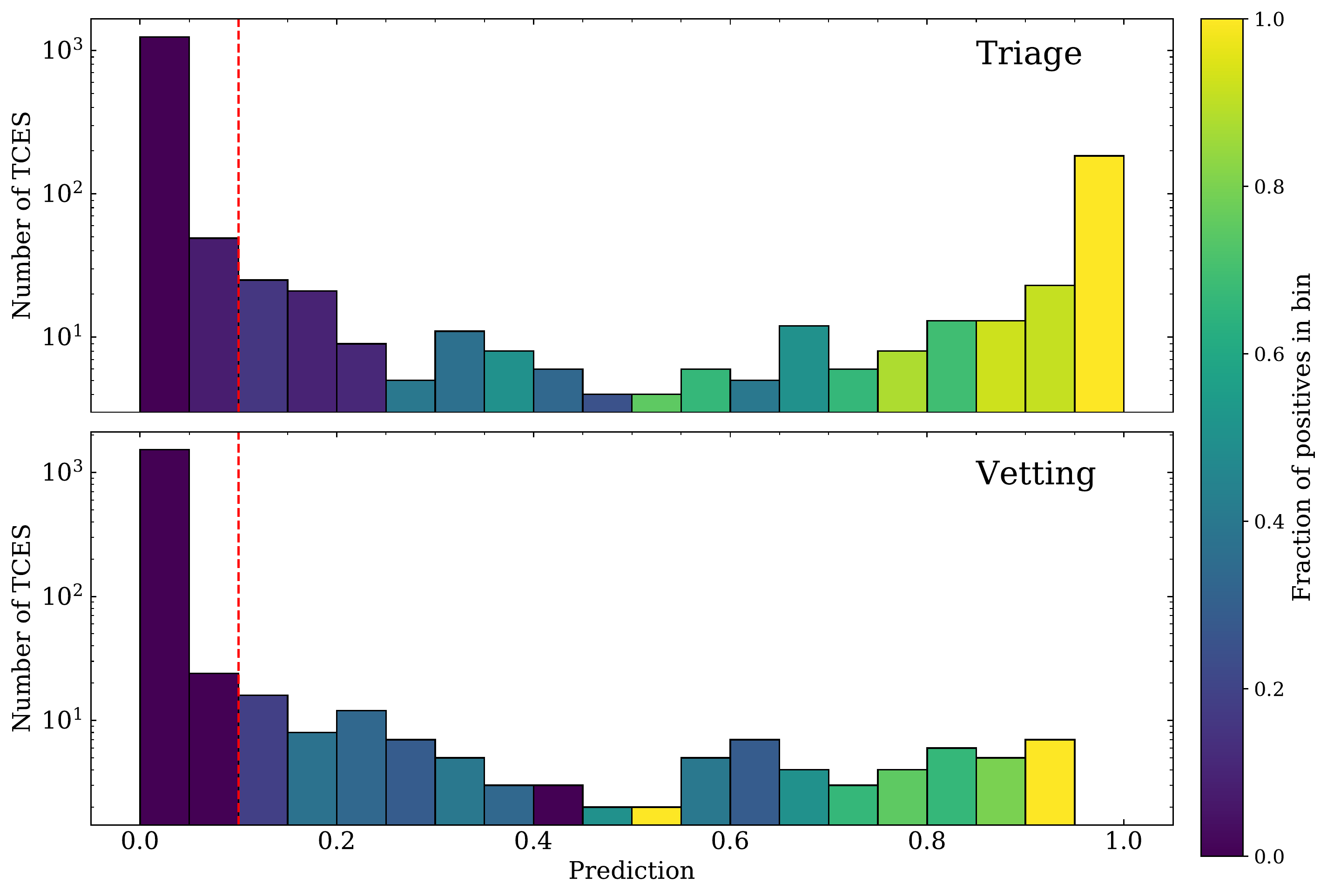}
    \caption{Histogram of predictions on the test set by \texttt{AstroNet-Triage} (top) and \texttt{AstroNet-Vetting} (bottom). The color of each bar represents the fraction of TCEs in that bin that are actually positives (PCs and EBs in triage mode, and just PCs in vetting mode). A yellow bin contains mostly TCEs that are positives, while a blue bin contains mostly negatives. The red dashed line marks a classification threshold of 0.1, which we find to maximize the fraction of false positives eliminated while still retaining almost all of the PCs.}
    \label{fig:hist}
\end{figure*}

We can also visualize the results in a different way. Fig.~\ref{fig:hist} shows a histogram of predictions given by the model to our test sets. The prediction loosely represents the probability that the model considers a given TCE to be a ``positive", meaning either a PC or EB in triage mode, or a PC in vetting mode. The color of each bar corresponds to the fraction of TCEs in that bin that are truly positives: a yellow bin contains mostly TCEs that are positives, while a blue bin contains mostly negatives. A good classifier would assign high predictions to positives and low predictions to negatives, so as to produce a histogram with yellow bins on the right side and blue bins on the left side. This is indeed what we see in both histograms.

\texttt{AstroNet-Triage} is already capable of achieving high precision and recall. Since the primary goal of triage is to cull the list of candidates while preserving most or all true PCs, we choose a classification threshold of 0.1 in order to discard only TCEs that the model is confident are false positives. With this classification threshold, we reach a precision of 0.749 and a recall of 0.975 on our test set of 1,650 TCEs. We recover all of the 49 PCs and the vast majority of EBs, while still eliminating 93\% of the negatives. 
The model can therefore be used to automatically eliminate obvious false positives in a set of TCEs with a minimal loss of PCs, allowing human vetters to focus instead on the strong candidate planets. The MIT \tess Team has already started using this model to perform triage on new TCEs from Sector 6 (see Section~\ref{sec:new}) and onward. 

\texttt{AstroNet-Vetting}, on the other hand, is less successful and not ready to be used in production. A natural classification threshold for vetting would be 0.5, which would select only those TCEs that the model considers more-likely-than-not planets. However, we find that the vetting model has difficulty distinguishing some PCs from EBs: at a threshold of 0.5, we recover just 28 of the 49 PCs from the same test set with a precision of 0.651. Since \tess is a mission designed with follow-up in mind, we would rather retrieve as many PCs as possible at the expense of more false positives, which can be easily vetted out by follow-up programs. We therefore choose to evaluate our vetting model at the more conservative threshold of 0.1.
At this threshold, we recover 44 of the 49 PCs with a precision of 0.449. Of the 5 missed PCs, three have systematics in their light curves that could have been mistaken for secondary eclipses, one is very V-shaped, and the last shows residual out-of-transit variability from imperfect detrending.
54 of the 69 false positives are EBs. A visual examination of the input representations of these misclassified EBs reveals that most do not have visible secondary eclipses nor exhibit significant changes in eclipse depth with aperture size. Most of these objects received EB labels during the initial inspection because they had odd-even transit differences or had synchronized out-of-transit variability that was later removed during detrending. These features are not captured in our input representation, so the model lacks sufficient information to distinguish these particular EBs from PCs. We discuss several ideas for improving the input representation in Section~\ref{sec:future}. The inability of the vetting model to separate EBs from PCs may also be due to the small number of PCs present in the training set. With the addition of new PCs from later \tess sectors, the model's performance in vetting mode may continue to improve. Still, our current results and success in triage mode indicate that our approach to automated vetting is a promising one. 

We note that even though \texttt{AstroNet-Vetting} cannot replace human vetting in its current state, and may never be able to do so completely, it can serve as a valuable complement to human vetting. This can help neutralize the shortcomings in both human and machine vetting. For example, it is difficult for human vetters to maintain a constant set of criteria when judging potential planet candidates, but machine learning can assign dispositions in a self-consistent, unbiased manner. On the other hand, a neural network can only detect patterns it was trained to detect. Unusual and interesting astrophysical signals that humans would recognize, such as WD 1145+017 b \citep{vanderburg15} and KIC 8462852 \citep{boyajian16}, would likely be classified as IS or V and discarded by neural networks. It would be useful to compare lists of PCs produced by humans and neural networks.

\section{Application to Previously Unseen TCEs}\label{sec:new}
\tess finished observing Sector 6 on Jan 7, 2019. We directly applied the trained \texttt{AstroNet-Triage} model to 59,719 new TCEs with the strongest BLS signals from Sector 6. Among these, 11,895 TCEs received a triage score of 0.1 or higher. We manually examined a random subset of 3,177 TCEs with triage scores of 0.1 or higher, and \tess magnitudes brighter than 12. Among these, we labeled 2,223 as EBs, 415 as PCs and 539 as IS or V. So if we accept these manually assigned labels as the ground truth, the precision of our model is 0.83 at a threshold of 0.1. Therefore our model is able to successfully eliminate a large number of false positives from Sector 6 TCEs, despite being trained on previous sectors that may have different systematics. It is worth noting that Sector 6 also covers a different stellar population compared to Sectors 1-5: because of its proximity to the Galactic plane, there are more evolved and variable stars in Sector 6. That our model was able to attain a precision comparable to that from Sectors 1-5 indicates that the model generalizes well to previously unseen sectors. We are also starting to see similar systematics from sector to sector now, so once we have built up a large sample from data taken using the same pointing strategy, we may achieve an even better performance when extrapolating to future sectors. 

\begin{figure*}[htb]
    \centering
    \includegraphics[width=\textwidth]{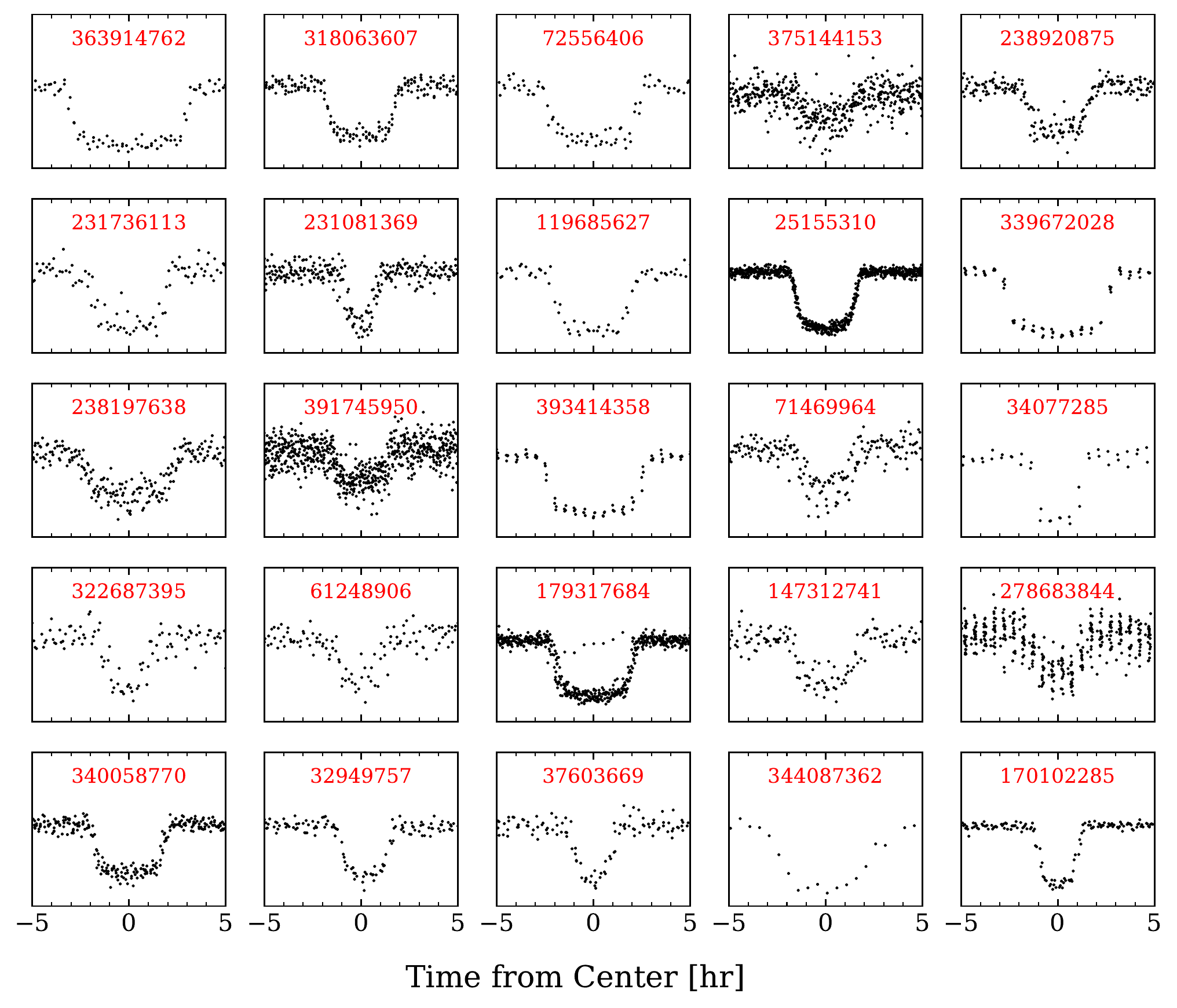}
    \caption{Phase-folded light curves of our 25 highest-quality planet candidates from Sector 6, along with their TIC IDs. To avoid clutter, we did not label the y-axis. Their transit parameters are listed in Table~\ref{tab:s6}.}
    \label{fig:pcs}
    \vspace{1em}
\end{figure*}

Although \texttt{AstroNet-Vetting} is not quite ready to be used in production, we generated scores for the manually examined subset of 3,177 TCEs with \texttt{AstroNet-Vetting} as a demonstration of what we can achieve with purely automated vetting at this stage. 700 of these TCEs received vetting predictions of 0.1 or higher, including 288 of the 415 PCs. Fig.~\ref{fig:pcs} shows 25 TCEs with the highest PC class probabilities that were also labeled as PCs by humans. At first glance, these do not show any warning signs of being non-planetary in nature (e.g. V-shaped transits or synchronized stellar activity). Our experience with \Kepler, \emph{K2} and earlier \tess sectors leads us to believe that most of these are indeed planetary in nature, and can quickly be confirmed via follow-up observations.  The transit properties from BLS for these TCEs and the remainder of the 273 highly ranked PCs are given in Table~\ref{tab:s6}. 

\section{Future Work}\label{sec:future}
\texttt{AstroNet-Triage} is already quite successful at distinguishing ``planet-like" TCEs (PCs and EBs) from instrumental noise and stellar variability, but both \texttt{AstroNet-Triage} and \texttt{AstroNet-Vetting} have room for improvement going forward. We have identified a few ways to improve these models in the future:

\begin{itemize}
    \item Currently, our training set only contains about $\sim 14,000$ TCEs, all of which are labeled by hand. It is therefore highly likely that there are some incorrectly labeled TCEs in the training set. Moreover, only $\sim 500$ of the TCEs are PCs. A larger, more accurately labeled data set would likely improve the performance of our model. Specifically, having more PCs on which to train should boost the accuracy of \texttt{AstroNet-Vetting}. One way to do this is to incorporate simulated transits injected into \tess light curves, but it is challenging to realistically simulate transit depth changes in different apertures, or to add simulated PCs with correct distributions of orbital periods and transit durations. If the simulated signals are sufficiently different from real PCs, including them may be detrimental to the model's performance. Future work may either explore how to accurately simulate TCEs, or retrain the model with new TCEs from future sectors.
    
    \item \citet{ansdell} showed that the inclusion of features such as stellar effective temperature, surface gravity, metallicity, radius, mass and density will likely improve our model's vetting accuracy. This is therefore a promising avenue for improving the model. In the future, we may amass a large enough sample of TCEs with stellar parameters from Gaia DR2 or the TIC to make this feasible.
    
    \item Including separate views of even- and odd-numbered transits may help \texttt{AstroNet-Vetting} identify eclipsing binaries with a true orbital period twice that reported by BLS and deep secondary eclipses present at phase 0.5.
    
    \item The interpolation method used to produce our input representations is not yet ideal. When generating binned views of light curves, we estimate the values of empty bins by linearly interpolating over neighboring bins (see Section~\ref{sec:rep}). This can distort the shapes of signals when there are large gaps in the data. A more intelligent interpolation method may be able to improve the model's performance. Alternatively, we could modify the model to take empty bins into account and avoid interpolation altogether.
    
\end{itemize}

Although our models currently only perform binary classification, they only require minor adjustments to perform multi-class classification. This may be of interest to researchers studying eclipsing binaries or stellar variability. 

\section*{Acknowledgments}
We thank Saul Rappaport and the anonymous referee for constructive feedback on the manuscript.

Funding for the \tess mission is provided by NASA's Science Mission directorate. 

A.V.'s work was performed under contract with the California Institute of Technology (Caltech)/Jet Propulsion Laboratory (JPL) funded by NASA through the Sagan Fellowship Program executed by the NASA Exoplanet Science Institute. Work by C.H. is supported by the Juan Carlos Torres Fellowship. I.J.M.C. acknowledges support from NSF through grant AST-1824644. T.D. acknowledges support from MIT’s Kavli Institute as a Kavli postdoctoral fellow. D.J.A. acknowledges support from the STFC via an Ernest Rutherford Fellowship (ST/R00384X/1).

{\it Facility:} \facility{\tess}


\newpage
\appendix
\LongTables
\begin{deluxetable*}{lrrllrll}[t]
\tabletypesize{\scriptsize}\tablecaption{New TCEs from Sector 6 with the highest likelihood of being planet candidates and manually assigned PC labels}\tablehead{
    TIC ID&  Tmag &      $P$ &       $T_0$ & Duration &  Depth  &                    Triage prediction & Vetting prediction                                         \\
    & & [d] & [BJD$_{TDB}$ - 2457000] & [hr] & [ppm] & &
}\startdata
363914762 & 10.931 & 13.862950 & 1445.332277 & 6.57 & 7490 & 0.894 & 0.935 \\
318063607 & 11.591 & 1.972508 & 1470.574615 & 3.98 & 6900 & 0.994 & 0.93 \\
72556406 & 10.763 & 5.564581 & 1470.665685 & 5.36 & 3590 & 0.976 & 0.926 \\
375144153 & 11.611 & 3.349557 & 1328.947672 & 3.20 & 1370 & 0.989 & 0.913 \\
238920875 & 11.740 & 6.534533 & 1326.067219 & 3.70 & 6610 & 0.935 & 0.911 \\
231736113 & 11.371 & 10.576018 & 1414.601978 & 4.52 & 5180 & 0.798 & 0.906 \\
231081369 & 11.686 & 7.632922 & 1329.841961 & 2.04 & 4880 & 0.968 & 0.906 \\
119685627 & 11.396 & 5.033517 & 1472.754875 & 4.71 & 9400 & 0.97 & 0.899 \\
25155310 & 10.555 & 3.288961 & 1327.516978 & 3.72 & 7020 & 0.996 & 0.894 \\
339672028 & 9.370 & 10.330855 & 1387.669219 & 5.89 & 4580 & 0.893 & 0.89 \\
238197638 & 11.729 & 7.276679 & 1355.883202 & 5.23 & 3270 & 0.945 & 0.89 \\
391745950 & 11.078 & 2.429812 & 1327.492932 & 3.07 & 1630 & 0.991 & 0.889 \\
393414358 & 10.417 & 4.374266 & 1469.700462 & 5.58 & 7210 & 0.994 & 0.886 \\
71469964 & 9.543 & 2.048315 & 1468.865706 & 3.15 & 900 & 0.978 & 0.878 \\
34077285 & 9.210 & 6.381659 & 1471.137677 & 2.93 & 3220 & 0.58 & 0.872 \\
322687395 & 11.415 & 4.002862 & 1471.141177 & 2.32 & 4300 & 0.967 & 0.871 \\
61248906 & 11.792 & 2.993511 & 1469.423987 & 2.64 & 3790 & 0.982 & 0.864 \\
179317684 & 10.843 & 4.231651 & 1328.874103 & 4.67 & 7190 & 0.992 & 0.862 \\
147312741 & 11.391 & 2.816972 & 1471.197891 & 3.57 & 3380 & 0.99 & 0.861 \\
278683844 & 9.234 & 5.542128 & 1327.600005 & 3.01 & 480 & 0.829 & 0.861 \\
340058770 & 11.874 & 2.758566 & 1385.912920 & 4.22 & 12690 & 0.996 & 0.857 \\
32949757 & 11.927 & 3.767558 & 1468.974521 & 2.86 & 11620 & 0.939 & 0.847 \\
37603669 & 11.584 & 2.969808 & 1468.864996 & 2.47 & 5020 & 0.99 & 0.845 \\
344087362 & 10.030 & 13.962886 & 1481.818801 & 6.84 & 3740 & 0.922 & 0.845 \\
170102285 & 11.682 & 2.941959 & 1470.667403 & 2.64 & 20280 & 0.982 & 0.844 \\
349789882 & 11.301 & 10.016470 & 1329.628043 & 1.69 & 1980 & 0.84 & 0.842 \\
443539530 & 11.158 & 2.719387 & 1470.030341 & 2.24 & 3290 & 0.97 & 0.832 \\
52640302 & 11.988 & 1.572030 & 1469.592407 & 2.55 & 16770 & 0.996 & 0.83 \\
235067594 & 11.276 & 8.296909 & 1438.933450 & 2.96 & 3750 & 0.892 & 0.826 \\
34371411 & 10.938 & 3.881647 & 1472.485586 & 4.54 & 8780 & 0.99 & 0.82 \\
172409263 & 9.995 & 2.111086 & 1469.560165 & 2.20 & 1140 & 0.974 & 0.816 \\
317924729 & 11.067 & 1.998197 & 1468.957994 & 3.61 & 11510 & 0.993 & 0.816 \\
255704097 & 10.585 & 6.014029 & 1470.980421 & 1.72 & 7740 & 0.86 & 0.815 \\
49079670 & 9.875 & 1.891807 & 1470.207475 & 1.44 & 770 & 0.887 & 0.809 \\
172464366 & 11.056 & 2.920137 & 1470.049317 & 3.17 & 14790 & 0.991 & 0.808 \\
443452168 & 11.857 & 4.634948 & 1472.707943 & 9.59 & 6880 & 0.986 & 0.808 \\
119170373 & 8.860 & 3.231364 & 1470.577387 & 1.65 & 1430 & 0.786 & 0.803 \\
25250808 & 11.515 & 3.323634 & 1468.782982 & 5.80 & 13550 & 0.995 & 0.803 \\
34466256 & 11.970 & 0.702749 & 1468.973241 & 1.20 & 2470 & 0.965 & 0.802 \\
150098860 & 9.656 & 10.692789 & 1335.921402 & 2.54 & 570 & 0.881 & 0.801 \\
322740947 & 11.883 & 1.750530 & 1470.221519 & 2.90 & 2120 & 0.987 & 0.799 \\
172193428 & 10.328 & 2.939634 & 1470.250746 & 1.25 & 1940 & 0.89 & 0.798 \\
38846515 & 10.307 & 2.849407 & 1326.744696 & 3.98 & 7730 & 0.998 & 0.797 \\
63571763 & 11.979 & 3.657340 & 1470.778793 & 3.04 & 3820 & 0.986 & 0.791 \\
79292541 & 9.373 & 2.275200 & 1470.742574 & 2.09 & 1490 & 0.946 & 0.79 \\
167418898 & 10.179 & 10.979537 & 1335.776524 & 1.92 & 2680 & 0.925 & 0.783 \\
157533118 & 11.733 & 2.519023 & 1470.833981 & 3.72 & 1510 & 0.782 & 0.776 \\
317483660 & 11.494 & 3.331287 & 1471.200359 & 4.10 & 11740 & 0.983 & 0.771 \\
54064834 & 11.495 & 6.057196 & 1470.396892 & 2.48 & 2960 & 0.92 & 0.768 \\
156836699 & 10.463 & 5.175336 & 1473.410078 & 3.05 & 3100 & 0.949 & 0.763 \\
201493205 & 10.619 & 4.063323 & 1472.533860 & 2.58 & 8260 & 0.97 & 0.757 \\
142523514 & 11.701 & 2.920137 & 1470.496760 & 2.34 & 7970 & 0.963 & 0.754 \\
339769761 & 11.442 & 4.604645 & 1386.570095 & 2.45 & 1520 & 0.949 & 0.752 \\
21725655 & 11.641 & 4.015693 & 1439.318766 & 2.39 & 3970 & 0.974 & 0.748 \\
67196573 & 10.729 & 2.556024 & 1469.400820 & 1.06 & 3280 & 0.689 & 0.734 \\
35644550 & 9.453 & 5.430301 & 1469.685013 & 3.83 & 2060 & 0.931 & 0.733 \\
139444326 & 11.474 & 3.526167 & 1440.702014 & 2.47 & 1300 & 0.641 & 0.732 \\
61404104 & 10.698 & 4.116773 & 1469.622464 & 3.64 & 1280 & 0.984 & 0.732 \\
63199675 & 10.425 & 2.833525 & 1470.244562 & 1.88 & 3530 & 0.954 & 0.723 \\
279644164 & 11.495 & 7.441509 & 1469.869921 & 4.49 & 15190 & 0.95 & 0.718 \\
200324182 & 10.342 & 1.297000 & 1411.750103 & 1.53 & 1890 & 0.991 & 0.716 \\
97279976 & 11.547 & 1.530326 & 1469.171958 & 3.44 & 5560 & 0.996 & 0.713 \\
172521714 & 11.279 & 3.831710 & 1470.174757 & 4.01 & 32160 & 0.99 & 0.712 \\
350445771 & 10.998 & 3.190404 & 1326.799310 & 1.57 & 2850 & 0.967 & 0.711 \\
317277995 & 11.873 & 2.433191 & 1470.987509 & 2.09 & 3610 & 0.92 & 0.708 \\
279425357 & 11.544 & 9.017454 & 1358.154457 & 1.44 & 4120 & 0.52 & 0.703 \\
279645722 & 10.369 & 2.220812 & 1469.322319 & 1.61 & 930 & 0.925 & 0.699 \\
78953309 & 9.725 & 1.931130 & 1470.346593 & 2.60 & 560 & 0.923 & 0.698 \\
38696105 & 10.482 & 5.577113 & 1326.104199 & 2.42 & 590 & 0.723 & 0.694 \\
232038804 & 11.217 & 4.154500 & 1471.244532 & 2.12 & 8930 & 0.96 & 0.693 \\
100589632 & 11.070 & 3.946770 & 1439.568240 & 1.81 & 1780 & 0.8 & 0.69 \\
192831602 & 11.381 & 9.790579 & 1447.827549 & 5.09 & 7200 & 0.927 & 0.686 \\
119544485 & 11.603 & 2.322156 & 1469.287125 & 2.93 & 1840 & 0.896 & 0.684 \\
63572800 & 11.829 & 2.021734 & 1470.356296 & 3.22 & 38350 & 0.991 & 0.674 \\
120165978 & 11.422 & 1.465310 & 1469.752267 & 3.06 & 2250 & 0.994 & 0.667 \\
24887574 & 11.001 & 1.771662 & 1468.921754 & 2.17 & 1020 & 0.718 & 0.66 \\
259701232 & 11.529 & 2.485701 & 1384.181439 & 3.18 & 4190 & 0.988 & 0.655 \\
443369587 & 11.885 & 2.085436 & 1470.142533 & 1.34 & 9360 & 0.965 & 0.652 \\
300116105 & 11.596 & 2.075595 & 1469.703864 & 1.42 & 3080 & 0.944 & 0.651 \\
346316941 & 11.412 & 1.688926 & 1469.289457 & 1.40 & 5510 & 0.918 & 0.651 \\
71728593 & 11.536 & 2.392378 & 1470.782565 & 2.53 & 13730 & 0.988 & 0.646 \\
443164624 & 11.764 & 2.525427 & 1469.482256 & 1.91 & 2620 & 0.921 & 0.641 \\
346574001 & 10.591 & 5.450762 & 1468.878060 & 3.64 & 820 & 0.699 & 0.641 \\
149603524 & 9.716 & 4.412208 & 1326.074373 & 3.95 & 14710 & 0.996 & 0.64 \\
200321330 & 11.143 & 1.815170 & 1411.405135 & 2.16 & 5130 & 0.987 & 0.635 \\
220459826 & 11.762 & 2.239526 & 1355.728967 & 1.26 & 1040 & 0.549 & 0.632 \\
124201045 & 11.899 & 7.157484 & 1474.437824 & 4.44 & 6760 & 0.85 & 0.631 \\
288078795 & 9.347 & 2.055383 & 1469.367685 & 1.68 & 2690 & 0.971 & 0.63 \\
147263084 & 11.743 & 0.614515 & 1468.868150 & 1.76 & 5660 & 0.956 & 0.624 \\
350623356 & 11.469 & 2.871340 & 1328.054642 & 2.66 & 660 & 0.944 & 0.618 \\
231969683 & 10.616 & 13.986153 & 1481.389627 & 3.54 & 13590 & 0.893 & 0.618 \\
238129783 & 11.277 & 4.849406 & 1469.394608 & 2.00 & 1680 & 0.3 & 0.612 \\
123742935 & 11.640 & 1.712055 & 1469.065779 & 1.81 & 2310 & 0.944 & 0.607 \\
47911178 & 9.776 & 3.586105 & 1470.300149 & 2.83 & 11850 & 0.98 & 0.605 \\
299742843 & 11.927 & 3.351131 & 1470.590447 & 1.89 & 8980 & 0.762 & 0.601 \\
140691463 & 11.976 & 2.084444 & 1326.551771 & 2.26 & 12920 & 0.992 & 0.6 \\
300146940 & 11.988 & 0.355657 & 1438.097487 & 1.31 & 5320 & 0.996 & 0.596 \\
63665162 & 11.783 & 3.985532 & 1471.517260 & 2.55 & 9520 & 0.969 & 0.59 \\
147977348 & 10.002 & 5.000113 & 1469.747290 & 3.51 & 6900 & 0.974 & 0.581 \\
123898871 & 9.831 & 4.901942 & 1470.359873 & 4.29 & 12330 & 0.98 & 0.576 \\
32925763 & 11.042 & 1.679960 & 1469.859626 & 1.45 & 2250 & 0.842 & 0.572 \\
306477840 & 10.981 & 5.522099 & 1469.524965 & 3.78 & 9700 & 0.939 & 0.57 \\
21725658 & 11.200 & 4.015693 & 1439.320057 & 2.48 & 2460 & 0.925 & 0.566 \\
48242396 & 11.709 & 0.865599 & 1468.736100 & 1.67 & 1760 & 0.943 & 0.566 \\
61341442 & 11.624 & 1.918690 & 1469.833807 & 3.79 & 2660 & 0.984 & 0.564 \\
339958786 & 11.703 & 7.497688 & 1389.519061 & 3.64 & 16280 & 0.954 & 0.561 \\
382626661 & 9.649 & 8.810778 & 1333.461016 & 3.79 & 280 & 0.712 & 0.553 \\
30031594 & 11.588 & 4.806822 & 1330.384295 & 2.70 & 1300 & 0.856 & 0.55 \\
142522973 & 11.821 & 7.097287 & 1473.393670 & 4.90 & 5790 & 0.969 & 0.54 \\
146918469 & 11.984 & 3.523713 & 1469.853577 & 3.34 & 8590 & 0.971 & 0.539 \\
200387965 & 11.673 & 0.550108 & 1411.443175 & 1.15 & 1300 & 0.979 & 0.536 \\
157568289 & 10.341 & 1.840512 & 1468.735923 & 5.17 & 3010 & 0.996 & 0.529 \\
443556801 & 11.266 & 1.508082 & 1470.123856 & 1.36 & 1550 & 0.7 & 0.523 \\
443115550 & 11.067 & 2.924354 & 1469.496132 & 3.25 & 1800 & 0.925 & 0.52 \\
97056348 & 11.956 & 2.898275 & 1471.068489 & 2.00 & 9330 & 0.949 & 0.518 \\
52452979 & 11.803 & 12.540751 & 1472.832850 & 4.19 & 5160 & 0.817 & 0.509 \\
32606889 & 11.585 & 4.684260 & 1440.937721 & 4.85 & 11120 & 0.987 & 0.502 \\
124331723 & 11.956 & 1.403180 & 1469.811385 & 3.58 & 10520 & 0.995 & 0.502 \\
35299896 & 11.809 & 7.057715 & 1470.284406 & 3.78 & 10120 & 0.734 & 0.499 \\
49187106 & 11.953 & 1.712634 & 1468.924649 & 3.14 & 1510 & 0.789 & 0.498 \\
14091704 & 9.136 & 0.764880 & 1438.420081 & 1.50 & 1900 & 0.995 & 0.493 \\
382101339 & 11.739 & 0.268842 & 1325.740785 & 0.74 & 630 & 0.895 & 0.492 \\
130613909 & 11.924 & 2.240024 & 1470.162214 & 1.79 & 11610 & 0.954 & 0.483 \\
349271454 & 11.575 & 0.716456 & 1325.793257 & 1.10 & 870 & 0.851 & 0.482 \\
238926217 & 11.983 & 3.351340 & 1326.984089 & 2.16 & 1370 & 0.948 & 0.481 \\
52639431 & 11.061 & 1.475015 & 1469.544384 & 2.18 & 3550 & 0.992 & 0.474 \\
78669071 & 11.280 & 1.516892 & 1469.596999 & 2.26 & 2990 & 0.988 & 0.469 \\
95418277 & 9.545 & 2.902560 & 1470.810932 & 3.41 & 460 & 0.808 & 0.465 \\
427352241 & 9.969 & 1.264720 & 1468.823180 & 2.31 & 2360 & 0.992 & 0.46 \\
33100834 & 11.332 & 5.741253 & 1473.985525 & 1.79 & 9060 & 0.841 & 0.459 \\
33797807 & 11.376 & 7.446982 & 1468.788619 & 4.27 & 1730 & 0.714 & 0.459 \\
35491505 & 11.987 & 2.714285 & 1471.002712 & 3.03 & 3080 & 0.979 & 0.458 \\
119024411 & 11.095 & 0.973156 & 1469.121692 & 1.93 & 5860 & 0.992 & 0.456 \\
232038798 & 11.275 & 4.154500 & 1471.244036 & 2.14 & 10810 & 0.902 & 0.456 \\
260268672 & 11.066 & 2.199328 & 1326.994207 & 1.31 & 550 & 0.81 & 0.452 \\
124493296 & 11.393 & 0.462865 & 1468.916825 & 1.65 & 3710 & 0.994 & 0.452 \\
46312336 & 11.418 & 4.592904 & 1439.292303 & 1.76 & 1730 & 0.813 & 0.447 \\
147478809 & 11.844 & 1.593514 & 1469.999134 & 2.70 & 3520 & 0.988 & 0.445 \\
10001673159 & 11.388 & 2.091282 & 1326.962632 & 5.19 & 690 & 0.888 & 0.441 \\
443129289 & 11.610 & 0.510670 & 1468.851587 & 1.56 & 3470 & 0.97 & 0.44 \\
14092291 & 11.802 & 1.908158 & 1439.077233 & 2.72 & 12650 & 0.993 & 0.435 \\
220397831 & 11.936 & 7.048540 & 1359.766935 & 11.43 & 810 & 0.924 & 0.434 \\
349576483 & 11.856 & 0.259746 & 1325.701037 & 0.99 & 740 & 0.981 & 0.433 \\
124106074 & 11.285 & 5.586068 & 1469.075132 & 4.38 & 6830 & 0.295 & 0.433 \\
157661381 & 11.533 & 0.911638 & 1469.024954 & 2.17 & 1620 & 0.98 & 0.428 \\
120544415 & 11.782 & 1.911447 & 1468.968402 & 3.25 & 4210 & 0.776 & 0.423 \\
142468550 & 11.850 & 6.658610 & 1469.047725 & 3.37 & 5190 & 0.959 & 0.421 \\
279322914 & 11.542 & 9.434735 & 1328.214438 & 6.12 & 21230 & 0.921 & 0.419 \\
134198986 & 11.648 & 1.012970 & 1468.802921 & 1.57 & 4530 & 0.933 & 0.417 \\
34366697 & 11.428 & 0.746141 & 1469.256947 & 1.83 & 5290 & 0.989 & 0.416 \\
72490088 & 11.895 & 0.944827 & 1469.220860 & 2.55 & 1520 & 0.918 & 0.413 \\
238082493 & 10.065 & 0.876523 & 1468.667915 & 1.91 & 1210 & 0.989 & 0.413 \\
72580791 & 11.379 & 1.754168 & 1469.274700 & 2.21 & 1690 & 0.945 & 0.411 \\
443115574 & 10.523 & 2.925198 & 1469.481622 & 2.73 & 1500 & 0.805 & 0.411 \\
461840150 & 11.424 & 0.538105 & 1468.549807 & 1.02 & 1620 & 0.931 & 0.408 \\
391745951 & 11.804 & 2.429695 & 1327.495315 & 3.14 & 1620 & 0.935 & 0.406 \\
20178111 & 10.244 & 1.734102 & 1468.622076 & 2.18 & 2290 & 0.994 & 0.406 \\
172308091 & 11.242 & 1.224808 & 1469.622611 & 2.57 & 1880 & 0.985 & 0.405 \\
333340702 & 11.292 & 2.029839 & 1469.033238 & 2.32 & 1450 & 0.906 & 0.405 \\
375090561 & 11.381 & 5.423940 & 1330.698602 & 3.15 & 2840 & 0.585 & 0.403 \\
339733013 & 10.038 & 5.620686 & 1328.386377 & 2.52 & 550 & 0.709 & 0.403 \\
63113815 & 10.432 & 1.738210 & 1469.892651 & 2.41 & 1590 & 0.986 & 0.401 \\
79142467 & 10.751 & 1.003933 & 1468.769629 & 3.30 & 1500 & 0.988 & 0.393 \\
157311499 & 11.142 & 2.010626 & 1470.124626 & 3.29 & 1050 & 0.744 & 0.392 \\
49669244 & 11.763 & 1.708877 & 1469.344553 & 4.68 & 9240 & 0.995 & 0.392 \\
30321299 & 11.105 & 3.864392 & 1327.954251 & 3.27 & 720 & 0.953 & 0.388 \\
350274840 & 11.624 & 1.597868 & 1326.048299 & 2.33 & 3230 & 0.99 & 0.386 \\
72090501 & 6.832 & 1.070700 & 1469.473875 & 2.16 & 3820 & 0.986 & 0.383 \\
49379306 & 11.418 & 2.116823 & 1469.224144 & 3.94 & 3460 & 0.983 & 0.382 \\
150437346 & 11.557 & 1.392659 & 1326.870639 & 2.05 & 7120 & 0.993 & 0.381 \\
79941130 & 11.975 & 1.698273 & 1469.433447 & 3.58 & 1920 & 0.903 & 0.38 \\
7420600 & 11.211 & 1.014010 & 1439.016749 & 2.30 & 2430 & 0.991 & 0.376 \\
238197709 & 10.260 & 6.864081 & 1354.344756 & 2.86 & 4030 & 0.918 & 0.374 \\
47711963 & 11.225 & 2.318965 & 1468.682246 & 1.68 & 1280 & 0.66 & 0.373 \\
63665158 & 11.785 & 3.987102 & 1471.515918 & 2.81 & 9270 & 0.944 & 0.366 \\
66915559 & 11.196 & 1.127067 & 1469.542826 & 3.45 & 3470 & 0.985 & 0.358 \\
34196883 & 11.631 & 1.617181 & 1469.404646 & 3.54 & 8310 & 0.996 & 0.355 \\
157129452 & 11.309 & 1.182573 & 1468.527976 & 3.32 & 1940 & 0.931 & 0.355 \\
35582553 & 9.840 & 0.935488 & 1468.837040 & 1.73 & 1360 & 0.98 & 0.348 \\
389920949 & 9.888 & 11.917522 & 1335.248973 & 3.96 & 5930 & 0.341 & 0.343 \\
278775625 & 11.215 & 5.128430 & 1328.941195 & 2.19 & 670 & 0.746 & 0.342 \\
421900585 & 11.449 & 6.903744 & 1472.660729 & 5.42 & 1990 & 0.873 & 0.339 \\
219151731 & 10.086 & 1.485150 & 1438.985678 & 2.37 & 1120 & 0.971 & 0.33 \\
443130801 & 10.576 & 2.169894 & 1470.664990 & 4.12 & 1270 & 0.942 & 0.326 \\
130415266 & 7.281 & 13.473506 & 1481.792518 & 6.39 & 8360 & 0.941 & 0.324 \\
35290793 & 10.970 & 0.280211 & 1468.827125 & 1.14 & 640 & 0.761 & 0.311 \\
52324253 & 10.318 & 1.676432 & 1469.319684 & 2.21 & 2510 & 0.992 & 0.307 \\
150186145 & 11.788 & 0.270739 & 1325.764847 & 0.93 & 2580 & 0.993 & 0.302 \\
157041282 & 11.995 & 6.491574 & 1474.795888 & 2.70 & 9010 & 0.805 & 0.302 \\
317548889 & 6.781 & 6.861642 & 1469.573795 & 3.75 & 230 & 0.488 & 0.301 \\
78820372 & 10.373 & 0.812993 & 1469.265872 & 1.37 & 800 & 0.874 & 0.293 \\
101144450 & 11.487 & 4.368986 & 1470.701150 & 2.17 & 1600 & 0.59 & 0.292 \\
348995211 & 11.333 & 0.345693 & 1325.753557 & 1.34 & 2260 & 0.996 & 0.29 \\
79682476 & 11.748 & 3.357798 & 1470.594351 & 4.10 & 1850 & 0.916 & 0.288 \\
219421728 & 11.154 & 0.671078 & 1411.370168 & 1.65 & 3950 & 0.996 & 0.287 \\
284288080 & 10.783 & 1.834544 & 1469.906630 & 2.99 & 970 & 0.593 & 0.284 \\
30538087 & 11.740 & 4.136415 & 1355.799976 & 5.80 & 830 & 0.713 & 0.284 \\
317022315 & 11.968 & 2.226672 & 1469.258933 & 4.81 & 4500 & 0.983 & 0.28 \\
124323593 & 11.305 & 5.589151 & 1469.077755 & 4.23 & 5160 & 0.249 & 0.279 \\
79139296 & 11.697 & 1.516437 & 1468.911918 & 2.44 & 1550 & 0.865 & 0.279 \\
157404343 & 8.352 & 3.139683 & 1470.492164 & 3.63 & 450 & 0.972 & 0.271 \\
172410994 & 11.392 & 0.453232 & 1468.691820 & 1.19 & 1490 & 0.795 & 0.268 \\
346488066 & 10.601 & 0.834499 & 1469.438340 & 1.95 & 840 & 0.9 & 0.267 \\
200326413 & 10.356 & 0.455446 & 1411.412194 & 1.07 & 900 & 0.975 & 0.266 \\
52812339 & 11.831 & 5.513609 & 1473.895818 & 3.90 & 2000 & 0.611 & 0.266 \\
32641207 & 11.536 & 0.407934 & 1438.509538 & 1.83 & 1710 & 0.992 & 0.265 \\
48752342 & 10.087 & 1.614603 & 1469.310972 & 2.33 & 5170 & 0.985 & 0.263 \\
35410741 & 11.047 & 1.049614 & 1469.351019 & 3.37 & 1130 & 0.954 & 0.262 \\
156992575 & 10.520 & 0.486501 & 1468.795743 & 2.05 & 1180 & 0.991 & 0.259 \\
52169698 & 10.885 & 0.622254 & 1468.530363 & 1.53 & 1300 & 0.979 & 0.257 \\
157566468 & 11.288 & 1.067850 & 1468.617298 & 2.32 & 6830 & 0.992 & 0.257 \\
443257841 & 11.943 & 2.697654 & 1471.092085 & 1.94 & 2700 & 0.543 & 0.257 \\
47773319 & 11.797 & 0.693984 & 1468.691629 & 1.72 & 8850 & 0.993 & 0.257 \\
231717034 & 10.771 & 2.198655 & 1384.483532 & 3.35 & 2440 & 0.229 & 0.248 \\
388128308 & 11.955 & 1.194126 & 1325.815270 & 2.09 & 8620 & 0.993 & 0.246 \\
31142436 & 11.714 & 5.271984 & 1440.303827 & 4.11 & 1080 & 0.453 & 0.245 \\
35488933 & 11.880 & 2.173153 & 1469.109265 & 2.52 & 1730 & 0.696 & 0.245 \\
388850377 & 11.094 & 2.467580 & 1469.295266 & 1.95 & 1040 & 0.577 & 0.245 \\
34377352 & 11.594 & 7.142339 & 1469.185327 & 3.51 & 4860 & 0.828 & 0.243 \\
287995512 & 11.938 & 0.938523 & 1468.930512 & 2.34 & 10560 & 0.965 & 0.242 \\
34521303 & 11.848 & 1.891453 & 1469.549945 & 3.47 & 1660 & 0.743 & 0.24 \\
220397824 & 11.379 & 7.049527 & 1359.742831 & 10.45 & 510 & 0.495 & 0.238 \\
37770169 & 10.650 & 6.097314 & 1474.488281 & 3.85 & 1860 & 0.835 & 0.238 \\
369517674 & 11.714 & 0.713896 & 1469.042390 & 2.15 & 1730 & 0.733 & 0.237 \\
32643071 & 10.570 & 2.161556 & 1470.560955 & 3.85 & 1090 & 0.972 & 0.234 \\
348995212 & 11.471 & 0.345693 & 1325.753438 & 1.34 & 2580 & 0.994 & 0.233 \\
34790951 & 11.363 & 4.974178 & 1473.246406 & 3.85 & 5310 & 0.579 & 0.227 \\
443451099 & 11.829 & 3.133377 & 1470.767509 & 3.70 & 6890 & 0.981 & 0.225 \\
120540763 & 11.908 & 2.261481 & 1470.124763 & 3.31 & 2900 & 0.672 & 0.225 \\
25413404 & 11.333 & 1.921603 & 1469.926743 & 4.07 & 4410 & 0.989 & 0.221 \\
123457307 & 11.995 & 2.870472 & 1469.458499 & 4.63 & 14430 & 0.974 & 0.217 \\
31852980 & 9.821 & 7.412709 & 1327.144736 & 4.09 & 350 & 0.203 & 0.214 \\
33602950 & 10.780 & 0.816532 & 1468.952009 & 2.70 & 770 & 0.46 & 0.213 \\
92845561 & 11.352 & 5.651537 & 1442.294869 & 3.08 & 1350 & 0.897 & 0.212 \\
255588086 & 10.869 & 0.896959 & 1438.655066 & 1.80 & 2760 & 0.992 & 0.211 \\
30848598 & 10.791 & 0.753911 & 1326.409307 & 3.29 & 570 & 0.974 & 0.211 \\
79143083 & 10.314 & 1.503826 & 1468.975437 & 3.09 & 3170 & 0.993 & 0.208 \\
120540056 & 11.158 & 1.522823 & 1469.819949 & 1.76 & 2560 & 0.959 & 0.207 \\
124022931 & 11.994 & 5.589151 & 1469.072801 & 4.63 & 19330 & 0.83 & 0.206 \\
48806546 & 11.074 & 0.932731 & 1468.802668 & 1.91 & 600 & 0.51 & 0.196 \\
219205407 & 10.804 & 6.125959 & 1327.329607 & 1.82 & 34050 & 0.977 & 0.187 \\
150066562 & 10.186 & 0.978050 & 1325.961256 & 1.11 & 420 & 0.404 & 0.185 \\
63423599 & 10.661 & 1.185797 & 1469.632930 & 4.90 & 4070 & 0.997 & 0.184 \\
147375101 & 10.741 & 1.586524 & 1469.185487 & 3.01 & 730 & 0.432 & 0.184 \\
299655932 & 11.433 & 1.331064 & 1468.785574 & 1.76 & 1750 & 0.77 & 0.179 \\
124097546 & 11.479 & 0.750065 & 1469.124751 & 1.85 & 1870 & 0.818 & 0.177 \\
238192097 & 11.919 & 1.227093 & 1325.744153 & 2.20 & 24100 & 0.996 & 0.177 \\
172409594 & 11.869 & 0.341709 & 1468.731897 & 1.33 & 2290 & 0.919 & 0.175 \\
349483495 & 11.643 & 0.998846 & 1326.031092 & 2.31 & 7380 & 0.997 & 0.173 \\
167714792 & 11.072 & 0.929267 & 1438.106543 & 1.58 & 1030 & 0.822 & 0.171 \\
382302241 & 10.976 & 1.598020 & 1326.041226 & 2.40 & 2010 & 0.98 & 0.168 \\
94989423 & 11.697 & 0.977006 & 1438.996675 & 1.11 & 1240 & 0.899 & 0.164 \\
260708537 & 9.342 & 1.744675 & 1326.979158 & 1.26 & 190 & 0.566 & 0.16 \\
333426440 & 11.379 & 2.423293 & 1470.386555 & 3.75 & 15780 & 0.99 & 0.157 \\
404965758 & 11.854 & 0.663947 & 1326.128108 & 3.48 & 2000 & 0.973 & 0.155 \\
143350974 & 11.608 & 1.081449 & 1439.142065 & 2.23 & 10260 & 0.98 & 0.154 \\
34443859 & 11.583 & 1.676900 & 1469.862528 & 2.61 & 1800 & 0.789 & 0.154 \\
349311188 & 11.291 & 5.608167 & 1326.174989 & 4.66 & 510 & 0.507 & 0.151 \\
340797848 & 11.702 & 7.387224 & 1474.903044 & 5.04 & 4030 & 0.669 & 0.15 \\
201508515 & 11.480 & 0.986094 & 1468.913865 & 1.71 & 1250 & 0.278 & 0.149 \\
48176862 & 11.412 & 1.925622 & 1469.928224 & 4.17 & 20480 & 0.993 & 0.146 \\
151628217 & 11.022 & 1.111059 & 1438.206509 & 2.38 & 9600 & 0.995 & 0.145 \\
92880568 & 10.924 & 0.588447 & 1438.538668 & 1.40 & 3470 & 0.993 & 0.144 \\
33002823 & 11.063 & 0.737931 & 1469.091985 & 1.96 & 1840 & 0.97 & 0.143 \\
55272169 & 11.385 & 1.008285 & 1326.298004 & 1.93 & 430 & 0.68 & 0.141 \\
260709785 & 11.896 & 1.156873 & 1325.754191 & 1.76 & 650 & 0.274 & 0.138 \\
53823382 & 11.671 & 3.452825 & 1470.817027 & 4.69 & 7590 & 0.984 & 0.135 \\
31109502 & 11.411 & 4.077800 & 1329.635600 & 3.44 & 520 & 0.731 & 0.134 \\
31054498 & 9.879 & 1.411197 & 1439.448140 & 3.06 & 520 & 0.949 & 0.132 \\
93123746 & 11.991 & 0.634459 & 1438.554379 & 3.70 & 1960 & 0.952 & 0.131 \\
32050278 & 10.889 & 9.040116 & 1325.996384 & 3.51 & 6510 & 0.864 & 0.131 \\
79439026 & 11.621 & 0.787574 & 1468.899673 & 2.00 & 4470 & 0.978 & 0.13 \\
201369213 & 11.508 & 2.809158 & 1469.713993 & 3.63 & 10470 & 0.961 & 0.125 \\
78672342 & 10.118 & 2.976792 & 1471.303365 & 3.44 & 910 & 0.872 & 0.123 \\
25191560 & 9.787 & 2.150539 & 1469.403285 & 3.83 & 910 & 0.856 & 0.121 \\
78956561 & 11.109 & 2.028213 & 1469.845112 & 4.02 & 930 & 0.321 & 0.12 \\
393159572 & 10.821 & 1.403214 & 1469.733819 & 2.57 & 2490 & 0.991 & 0.119 \\
123958679 & 11.426 & 5.589151 & 1469.074462 & 4.14 & 4210 & 0.164 & 0.119 \\
238006656 & 11.314 & 0.877241 & 1354.366910 & 1.73 & 880 & 0.935 & 0.116 \\
71917644 & 11.394 & 1.109653 & 1468.838374 & 2.67 & 1680 & 0.798 & 0.115 \\
364395234 & 11.708 & 1.375841 & 1326.602612 & 2.70 & 7660 & 0.998 & 0.114 \\
124543547 & 10.486 & 5.355631 & 1471.746463 & 7.07 & 3300 & 0.866 & 0.114 \\
120027834 & 11.518 & 2.996169 & 1471.466335 & 3.66 & 2570 & 0.6 & 0.111 \\
375032908 & 9.328 & 8.518659 & 1333.616043 & 3.21 & 530 & 0.26 & 0.11 \\
143218704 & 11.692 & 1.857112 & 1469.812388 & 5.55 & 30850 & 0.997 & 0.106 \\
317876382 & 10.966 & 2.150539 & 1470.621890 & 4.18 & 950 & 0.774 & 0.104 \\
382068562 & 11.293 & 12.129942 & 1330.347472 & 2.15 & 16680 & 0.959 & 0.102 \\
4616346 & 11.498 & 0.690059 & 1468.773630 & 1.77 & 1230 & 0.409 & 0.1
\enddata\label{tab:s6}
\end{deluxetable*}

\end{document}